 \documentclass[sn-mathphys-num]{sn-jnl}

\usepackage{graphicx}%
\usepackage{multirow}%
\usepackage{amsmath,amssymb,amsfonts}%
\usepackage{amsthm}%
\usepackage{mathrsfs}%
\usepackage[title]{appendix}%
\usepackage{xcolor}%
\usepackage{textcomp}%
\usepackage{manyfoot}%
\usepackage{booktabs}%
\usepackage{algorithm}%
\usepackage{algorithmicx}%
\usepackage{algpseudocode}%
\usepackage{listings}%

	\def\beq{\begin{equation}}
	\def\endeq{\end{equation}}
	\def\begdi{\begin{displaymath}}
	\def\enddi{\end{displaymath}}
	
	\def\halfpi{\frac{\pi}{2}}
			\large

    \def\FI{${\mathcal F}_{\rm I}$\ }
    \def\FII{${\mathcal F}_{\rm II}$\ }
    \def\FIII{${\mathcal F}_{\rm III}$\ }
    \def\FB{${\mathcal F}_{\mathcal B}$\ }	
    \def\B{{\mathcal {B}}}
    \def\I{{[\rm I]}}
    \def\II{{[\rm II]}}
    \def\III{{[\rm III]}}

\begin{document}

\title[Article Title]{Extension of the creep tide theory to exoplanet systems with high stellar obliquity. The dynamic tide of CoRoT-3b}

\author[1]{\fnm{Hugo} \sur{Folonier}}\email{hugofolonier@gmail.com}

\author[2]{\fnm{Sylvio} \sur{Ferraz-Mello}}\email{sylvio@iag.usp.br}

\author[2]{\fnm{Raphael} \sur{Alves-Silva}}\email{alves.raphael@usp.br}

\affil[1]{\orgname{Observatorio Astron\'omico de C\'ordoba}, \city{C\'ordoba}, \country{Argentina}}

\affil[2]{\orgdiv{Instituto de Astronomia, Geof\'isica e Ci\^encias Atmosf\'ericas}, \orgname{Universidade de S\~ao Paulo}, \city{S\~ao Paulo}, \country{Brazil}}



\abstract{This paper extends the creep tide theory to exoplanetary systems with significant obliquities. The extended theory allows us to obtain the stellar and planetary hydrodynamic equilibrium tides and the evolution of the rotational state of the bodies. The dynamic ellipsoidal figure of equilibrium of the body is calculated taking into account that its reaction to external forces is delayed by its viscosity. The derived equations are used to determine the motion of the tidal bulge of the planetary companion CoRoT-3b (a brown dwarf) and its host star. We show how the tides deform the figure of the companion and how its tidal bulge moves close to the substellar meridian from one hemisphere to another. The stellar lag is mostly positive and is braking the star rotation.}





\maketitle

\section{Introduction}\label{sec1}
This paper deals with extending the creep tide theory (Ferraz-Mello 2012, 2013, 2015) to systems with significant obliquity.
These three-dimensional configurations occur when the rotation axis of the tidally deformed body (the primary) is not perpendicular to the orbital plane of the companion. 
This paper is a sequel to Folonier et al. (2022), which considered the tidal deformations of a perfect rotating body without viscosity (the so-called static tide) and presented the figure of equilibrium of the body under the joint action of both gravity and rotation (see Sect. \ref{sec:static}).
Now, the body's viscosity is considered, and the body's response to the tidal forces is assumed to be that of a low-Reynolds-number flow. The radial displacements follow a Newtonian creep law. 
This physical law and ordinary Physics are sufficient to construct the entire theory without the need for additional hypotheses about the body's deformation.

The creep tide theory has been used in many solar and extrasolar planetary systems studies. It allowed us to show that in a close-in extra-solar system with solar-type stars and large companions (hot Jupiters or brown dwarfs), the transfer of angular momentum from the planetary orbit via tides compensates for the leakage of angular momentum due to the stellar wind braking. 
Various examples with good-quality observational data were studied, and it was shown that the present stellar rotation is consistent with the angular momentum transfer and the known age of the system (Ferraz-Mello et al. 2015).
At variance, close-in systems with large companions hosted by F-stars showed synchronization of the stellar rotation and the planet's orbital motion. 
The creep tide theory has also been used in studies designed to understand the accumulation of short-period hot Jupiters around solar-type stars with rotation periods close to 25 days and the dearth of close-in planets around fast rotators (Ferraz-Mello and Beaug\'e, 2023).

In the applications to the Solar System, the best results were obtained in the study of Enceladus and Mimas (Folonier et al. 2018, 2019) and they were consistent with the very different heat dissipations in these two satellites, as well as with the observed forced libration of Enceladus, with no additional hypotheses. It was also used in the study of Mercury (Ferraz-Mello, 2015; Gomes et al. 2019). In this case, it unveiled a completely new mechanism of capture into the 3/2 resonance, totally different from the classical model based on the spin-orbit dynamics of rigid bodies (Ferraz-Mello, 2015). However, we must keep in mind that the current values for Mercury's equatorial prolateness and polar oblateness are 2-3 orders of magnitude larger than the values given by the tidal theories (see Gomes et al. 2019) making clear the need for more complex models allowing the primary's crust to remain solid (e.g. Bou\'e et al., 2017).  

All studies mentioned above were performed using the creep tide theory for systems with coplanar equators and orbit. However, as the information about the discovered exoplanets gradually increases, we start learning that there are many systems in which the orbit is inclined with respect to the star's equator (see Albrecht et al. 2022). Many of them are 2-body systems where the companion is a close-in hot Jupiter or brown dwarf in which tides may play an important role and for which the existing theory for systems with coplanar equators and orbit cannot be used. 

This paper extends the creep tide theory to the full three-dimensional case. It is complete in the sense that the full equations for the orbital and rotational elements and the figure of the bodies are given. It is derived from previous results on the full three-dimensional static tide theory developed by Folonier et al. (2022). Thus, it takes into account that the tidal bulges oscillate about the plane of the orbit instead of being fixed, as considered in most of the 3D tidal theories. However, it has an important limitation. By considering simultaneously the very short period variations of the figure of the bodies, it gives rise to a stiff system of 22 differential equations that can only be used to investigate the evolution of the systems for short times. In the case of extrasolar planetary systems, this limitation is of the order of a few years because of the relatively high value of the relaxation factor of stars and hot Jupiters. For the study of long-term evolution, the corresponding averaged equations are needed. However, the averaged equations erase the short-term variations considered in this paper and cannot be used for their study. They are treated separately (paper in preparation). 

As some of the most recent tidal theories (Correia et al. 2014, Bou\'e et al. 2016, Folonier et al. 2018, Ferraz-Mello et al. 2022), the developments here are based on a set of parametric equations fixing the figure and orientation of the tidally deformed body (dynamic tide or dynamic equilibrium tide)\footnote{{As classically defined in mechanics, dynamic equilibrium means that the forces applied to a system of massive particles and the time derivatives of the momenta of the system itself projected onto any virtual displacement are equal (d'Alembert). It differs from the static equilibrium where the projected applied forces are equal to zero. It also differs from the dynamical tide introduced by Zahn, which considers the dissipation due to the internal oscillation modes excited directly by the tidal forcing (see Ogilvie, 2014)}}: the equatorial protrusion, the polar oblateness, the lag of the bulge with respect to the sub-companion meridian and the latitude of the bulge with respect to the equator. The parametric equations are 5 equations, but numerical integration of the parametric equations showed that they are not fully independent. A relationship between them seems to exist. This relationship would extend to the dynamic tide a property previously found in the static tide according to which the axis of rotation is located in the plane defined by the vertex of the tidal bulge and the pole of the figure (Folonier et al. 2022). 

The theory developed in this work is applied to the planetary system of CoRoT-3. The stellar obliquity (inclination of the orbital plane over the equator of the star) has been measured by Triaud et al. (2009) as $37.6^{+10}_{-22.3}$ degrees. The planetary obliquity is not known, but to explore the outcomes in such a case, an arbitrary value was assumed ($\sim 18$ degrees). 

The theory is complex with several parts intertwining, making a linear presentation impossible. For ease of reading, the content was grouped into several parts. In the first part (Sections \ref{sec:figure} -- \ref{sec:I}), we present the parameters characterizing the static and dynamic tide (the Ariadne thread of this paper). In the second part (Sections \ref{sec:Pot} -- \ref{sec11}), we present the dynamics, and, in the third part (sections \ref{Part4} -- \ref{sec:longperiod}), we present an application of the theory to the exoplanetary system formed by the star CoRoT-3 and the brown dwarf CoRoT-3b, with results on the dynamic equilibrium tide of the two bodies.  The results show some interesting features. The tidal bulges oscillate in latitude. The stellar lag in longitude remains positive (as expected) most of the time. For the sake of comparison, the theory was also applied to a clone system moving around a slow-rotating star. The results are similar with the stellar lag now remaining negative.

 In the appendix, we separately present the four reference systems used in the study of dynamic tide in the body defined as the primary.
They are the astrocentric reference frame with inertial axes \FI; the equatorial reference frame \FII; the static equilibrium reference frame \FIII; and the body reference frame \FB. A glossary of the quantities related to the frame transformations is also given.

\section{The statement of the problem}\label{sec2}
Let us consider a system formed by a rotating fluid body $\mathsf{m}$ of mass $m$ (the primary), and a point mass $\mathsf{M}$ of mass $M$ (the secondary, also referred to as companion), and let ${\bf r}$ be the radius vector of $\mathsf{M}$ in a reference system centered on the primary $\mathsf{m}$. In the present work, we treat the case in which the bodies are homogeneous and the angular velocity vector (or spin) of the primary, ${\bf \Omega}$, is not perpendicular to the orbital plane. $\theta$ is the angle between the spin vector ${\bf \Omega}$ and the radius vector ${\bf r}$.

In the creep tide theory, the tidal deformation of $\mathsf{m}$ is assumed to be a low-Reynolds number flow  and is given by the Newtonian creep law
\begin{equation}
	\dot{\zeta}=\gamma(\rho-\zeta),
\end{equation}
where $\zeta=\zeta(\varphi_s,\theta_s,t)$ is the distance to the center of a surface point of spherical coordinates ${\varphi}_s,{\theta}_s$, respectively the longitude and the co-latitude of that point. Both are defined with respect to a reference system rotating with the primary body $\mathsf{m}$, with the origin centered at the center of mass of $\mathsf{m}$. 
$\rho=\rho({\varphi}_s,{\theta}_s,t)$ is the static equilibrium figure of the primary, that is, the equilibrium figure that it would assume if it were a perfect fluid (of zero viscosity), under the actions of its rotation and the gravitational perturbation from the companion $\mathsf{M}$.

The proportionality coefficient $\gamma$ is the relaxation factor.
The solution of the corresponding Navier-Stokes equations shows that it is given by:
\begin{equation}
	\gamma = \frac{dgR}{2\eta},
\end{equation}
where $d$, $R$, and $\eta$ are the density, the mean radius, and the viscosity of the primary, respectively, and $g$ is the gravity acceleration at the surface of $\mathsf{m}$ (see Ferraz-Mello, 2013; Folonier et al. 2018).

It is worth emphasizing that no hypothesis is made about the mass ratio $M/m$. In one system star-exoplanet, for example, the primary may be either the star or the planet. In fact, in a complete study, both cases need to be considered and their effects shall be added to get the complete tidal perturbation on the orbit of the planet around the star.

\section{The equilibrium figure}\label{sec:figure}
	
Any well-behaved function of longitudes and latitudes may be represented by a series of spherical harmonics. If additionally, we introduce symmetry with respect to the center, the lower-order terms correspond to triaxial ellipsoids. In the present theory, two of these approximations will be used. They are the static tide and the dynamic tide ellipsoidal figures of equilibrium. 

\subsection{Static tide ellipsoidal figure of equilibrium}\label{sec:static}

\begin{figure}[h]
	\begin{center}
			\includegraphics[width=11cm]{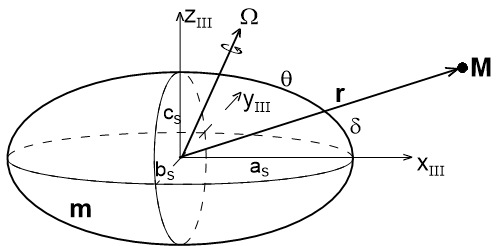}
	\end{center}
	\caption{Static tide ellipsoidal figure of equilibrium of the primary body $\mathsf{m}$ and reference system \FIII defined by the ellipsoid axes of symmetry  (see Appendix \ref{sec:FIII}).  The major and minor axes of the ellipsoid are in the plane defined by the instantaneous rotational velocity of the body ($\mathbf \Omega$) and the radius vector of the companion ($\mathbf r$) (Folonier et al. 2022).}   
	\label{fig:H-2}
\end{figure}

The static tide figure of equilibrium is the figure of equilibrium of a rotating inviscid body submitted to external gravitational forces (Folonier et al. 2022).  It can be approximated by a triaxial ellipsoid with semi-major axes $a_s>b_s>c_s$. The radius vector $\rho$ of one point on the surface of the figure is completely determined if the angular velocity vector ${\bf \Omega}$ and the radius vector $\mathbf{r}$ of the companion $\mathsf{M}$ are known.

As seen in Fig. \ref{fig:H-2}, the major axis vertex does not point towards the companion but lies at a distance $\delta$ from it. The semi-axes $a_s$ and $c_s$ lie in the plane defined by the radius vector ${\bf r}$ of the companion that attracts the body and the body's spin vector ${\bf \Omega}$.

Let us define the equatorial flattening and the polar oblateness of the static tide ellipsoid:
\begin{equation}
	\epsilon_\rho = \frac{a_s-b_s}{\sqrt{a_sb_s}}; \qquad \epsilon_z=1-\frac{c_s}{\sqrt{a_sb_s}}.
\end{equation}
We define $R=\sqrt[3]{a_sb_sc_s}$ as the mean radius of the primary, and to the first order in the flattenings, the semi-axes of the ellipsoid can be written as
\begin{equation}\label{eq:semiaxes-homo}
	a_s = R\left(1+\frac{\epsilon_\rho}{2}+\frac{\epsilon_z}{3}\right);  \quad  
	b_s = R\left(1-\frac{\epsilon_\rho}{2}+\frac{\epsilon_z}{3}\right);  \quad  
	c_s = R\left(1-\frac{2\epsilon_z}{3}\right).
\end{equation}

Also, from Folonier et al. (2022), it follows that the equatorial  flattening and the polar oblateness are
\begin{eqnarray} \label{flattenings}
	\epsilon_\rho &=& \frac{\epsilon_J-\epsilon_M}{2}+\frac{1}{2}\sqrt{\epsilon_J^2+\epsilon_M^2-2\epsilon_J\epsilon_M\cos{2\theta}},\nonumber\\
	\epsilon_z    &=& \frac{\epsilon_M-\epsilon_J}{4}+\frac{3}{4}\sqrt{\epsilon_J^2+\epsilon_M^2-2\epsilon_J\epsilon_M\cos{2\theta}},\label{eq:achata3}
\end{eqnarray}
where $\epsilon_J$ and $\epsilon_M$ are, respectively, the flattenings of the equivalent Jeans and  MacLaurin homogeneous ellipsoids (see Ferraz-Mello 2013, Ferraz-Mello et al. 2020), defined as follows:
\begin{equation} \label{def:em}
	\epsilon_J = \frac{15}{4}\left(\frac{M}{m}\right)\left(\frac{R}{r}\right)^3; \qquad \epsilon_M = \frac{5R^3\Omega^2}{4Gm}.
\end{equation}
The orientation angle $\delta$ between the major axis of the ellipsoid and  the direction of the companion satisfy the relations (Folonier et al. 2022):
\begin{eqnarray}
	\cos{2\delta}&=&\frac{\epsilon_J-\epsilon_M\cos{2\theta}}{\sqrt{\epsilon_J^2+\epsilon_M^2
			-2\epsilon_J\epsilon_M\cos{2\theta}}},\nonumber\\
	\sin{2\delta}&=&\frac{\epsilon_M\sin{2\theta}}{\sqrt{\epsilon_J^2+\epsilon_M^2
			-2\epsilon_J\epsilon_M\cos{2\theta}}}.
	\label{eq:delta}
\end{eqnarray}

In the previous equations, the angle $\theta$ is the co-latitude of the companion in the equatorial frame (\FII) defined as
\begin{equation}
	\theta = \cos^{-1}{\left(\frac{{\bf r}\cdot{\bf \Omega}}{r\Omega}\right)},
\end{equation}
where $r = |{\bf r}|$ and $\Omega = |{\bf \Omega}|$.

\subsection{The static tide parameters $f_i$}\label{sec:f_i}

Let us consider one generic point on the surface of the ellipsoid. In the equatorial frame $\mathcal{F}_{\rm II}$, its rectangular coordinates may be written:
\begin{eqnarray}\label{eq:coordinate-spherical-rho}
	x_{\II} &=& \rho \sin\theta_s \cos\varphi_s, \nonumber\\
	y_{\II} &=& \rho \sin\theta_s \sin\varphi_s, \\
	z_{\II} &=& \rho \cos\theta_s, \nonumber
\end{eqnarray}
where $\rho$ is the radius vector of the point and ($\varphi_s,\theta_s$) are its longitude and co-latitude in the frame $\mathcal{F}_{\rm II}$, respectively\footnote{The spherical coordinates (${\varphi_s},{\theta_s}$) are reckoned from the ascending node $\mathsf{N'}$ and the rotation pole, respectively (see Fig. \ref{fig:H-4} in the Appendix).}.

To obtain the coordinates of this point in the frame $\mathcal{F}_{\rm III}$, we may use the transformation introduced in the Appendix \ref{sec:FIII}:
\begin{equation}
\mathbf{v}_{[\rm III]}= \mathsf{P}^{-1}\mathbf{v}_{[\rm II]},
\end{equation}
that is,
\begin{eqnarray}
	x_{\III} &=&  x_{\II}\sin(\theta+\delta)\cos{\varphi} + 		
					   y_{\II}\sin(\theta+\delta)\sin{\varphi} + 
					   z_{\II}\cos(\theta+\delta), \nonumber\\
	y_{\III} &=& -x_{\II}\sin{\varphi} + 
					   y_{\II}\cos{\varphi}, \\
	z_{\III} &=& -x_{\II}\cos(\theta+\delta)\cos{\varphi}   
	                  -y_{\II}\cos(\theta+\delta)\sin{\varphi}
	                  +z_{\II}\sin(\theta+\delta), \nonumber
\end{eqnarray}
where $\varphi,\theta$ are the longitude and co-latitude of the companion in the frame $\mathcal{F}_{\rm II}$, respectively. 

The equation of the static tide equilibrium figure may be written as
\begin{equation}
	 \frac{x_{{\III}}^{2}}{a_s^2}+\frac{y_{{\III}}^2}{b_s^2}+\frac{z_{{\III}}^{2}}{c_s^2}-1=0,
\end{equation}
where $a_s,b_s,c_s$ are the semi-axes of the ellipsoid given by Eqn. (\ref{eq:semiaxes-homo}), 
to the first order in the flattenings $\epsilon_{\rho}, \epsilon_{z}$.
This equation can be solved with respect to $\rho$ giving
\begin{eqnarray}	\label{eq:rhoII}
	\rho &=& R\Big[1+f_1\mathcal{P}+f_2\mathcal{C}_2+f_3\mathcal{S}_2+f_4\mathcal{C}_1+f_5\mathcal{S}_1\Big].
\end{eqnarray}
where $\mathcal{P},\mathcal{C}_2,\mathcal{S}_2,\mathcal{C}_1,\mathcal{S}_1$ are functions of the angular coordinates of the surface point defined as
\begin{eqnarray}	\label{eq:PCS}
	\nonumber \mathcal{P}   &=& \frac{1}{3}-\cos^2\theta_s, \\
	\nonumber \mathcal{C}_1 &=& \sin2\theta_s\cos\varphi_s, \\
	\mathcal{S}_1 &=& \sin2\theta_s\sin\varphi_s, \\ 
	\nonumber \mathcal{C}_2 &=& \sin^2\theta_s\cos2\varphi_s,\\
	\nonumber \mathcal{S}_2 &=& \sin^2\theta_s\sin2\varphi_s, 
\end{eqnarray}
and the coefficients $f_i$ are functions of the static tide ellipsoid flattenings $\epsilon_\rho,\epsilon_\rho$, the orientation angle $\delta$, and the angular coordinates $\varphi,\theta$ of the companion:
\begin{eqnarray}	\label{eq:fi}
	f_1 &=& \epsilon_z-\frac{3}{2}\Big(\epsilon_z+\frac{\epsilon_\rho}{2}\Big)\cos^2{(\theta+\delta)},\nonumber\\
	f_2 &=& \Big[\frac{\epsilon_\rho}{2}-\frac{1}{2}\Big(\epsilon_z+\frac{\epsilon_\rho}{2}\Big)\cos^2{(\theta+\delta)}\Big]\cos{2\varphi},\nonumber\\
	f_3 &=& \Big[\frac{\epsilon_\rho}{2}-\frac{1}{2}\Big(\epsilon_z+\frac{\epsilon_\rho}{2}\Big)\cos^2{(\theta+\delta)}\Big]\sin{2\varphi}, \nonumber\\
	f_4 &=& \frac{1}{2}\Big(\epsilon_z+\frac{\epsilon_\rho}{2}\Big)\sin{(2\theta+2\delta)}\cos{\varphi},\nonumber\\
	f_5 &=& \frac{1}{2}\Big(\epsilon_z+\frac{\epsilon_\rho}{2}\Big)\sin{(2\theta+2\delta)}\sin{\varphi},  \nonumber
\end{eqnarray}
or, equivalently,
\begin{eqnarray}	\label{eq:fequiv}
	f_1 &=& \epsilon_M +\frac{1}{2}\epsilon_J (1-3\cos^2\theta),\nonumber\\
	f_2 &=& \frac{1}{2}\epsilon_J\sin^2\theta \cos{2\varphi},\nonumber\\
	f_3 &=& \frac{1}{2}\epsilon_J\sin^2\theta \sin{2\varphi},\\
	f_4 &=& \frac{1}{2}\epsilon_J\sin {2 \theta} \cos{\varphi},\nonumber\\
	f_5 &=& \frac{1}{2}\epsilon_J\sin {2\theta} \sin{\varphi},\nonumber
\end{eqnarray}

\subsection{Dynamic tide ellipsoidal figure of equilibrium}\label{sec:dynamic}

The dynamic tide ellipsoidal figure of equilibrium is the actual figure of the body considering that its reaction to the external forces is delayed by its viscosity. 
Its ellipsoidal approximation is the target of the present paper. 

Let us define the equatorial flattenings and polar oblateness of the dynamic tide ellipsoidal figure of equilibrium as:
\begin{equation}
	\mathcal{E}_\rho = \frac{a_\B-b_\B}{\sqrt{a_\B b_\B}}; \qquad \mathcal{E}_z=1-\frac{c_\B}{\sqrt{a_\B b_\B}}.
\end{equation}
We define $R=\sqrt[3]{a_\B b_\B c_\B}$ as the mean radius of the primary, and to the first order in the flattenings, the semi-axes of the ellipsoid can be written as
\begin{equation}
	a_\B = R\left(1+\frac{\mathcal{E}_\rho}{2}+\frac{\mathcal{E}_z}{3}\right);  \quad  b_\B = R\left(1-\frac{\mathcal{E}_\rho}{2}+\frac{\mathcal{E}_z}{3}\right);  \quad  c_\B = R\left(1-\frac{2\mathcal{E}_z}{3}\right).
	\label{eq:semiaxesdyn}
\end{equation}

\subsection{The dynamic tide parameters $E_i$}\label{sec:E_i}

Let us consider one generic point in the surface of the primary defined by the dynamic tide ellipsoid. {The equation of the ellipsoid is}
\begin{equation}
	\frac{x^{2}_{\B}}{a_\mathcal{B}^2}+
	\frac{y^{2}_{\B}}{b_\mathcal{B}^2}+
	\frac{z^{2}_{\B}}{c_\B^2}-1=0, \label{eq:elipsoide}
\end{equation}
where $x_\B, y_\B, z_\B$  are its coordinates in the frame $\mathcal{F}_\mathcal{B}$ defined by the principal axes of the ellipsoid (see Appendix  \ref{sec:FB}). The coordinates of one vector ${\bf v}$ in the $\mathcal{F}_\mathcal{B}$-frame can be written in terms of the coordinates in the equatorial frame  $\mathcal{F}_{{\rm II}}$, using the relation 
	\begin{equation}
		\mathbf{v}_{\mathcal{B}}= \mathsf{Q}^{-1}\mathbf{v}_{[\rm II]},
	\end{equation}
{and the coordinates}	
\begin{eqnarray}	\label{eq:coordinate-spherical-zeta}
	x_\II &=& \zeta \sin\theta_s \cos\varphi_s, \nonumber\\
	y_\II &=& \zeta \sin\theta_s \sin\varphi_s, \\
	z_\II &=& \zeta \cos\theta_s,\nonumber
\end{eqnarray}
where $\zeta$ is the radius vector of the point in the generic direction whose longitude and co-latitude, in the frame \FII, are ($\varphi_s,\theta_s$). 

Thus, we have

\begin{eqnarray}
	x_\mathcal{B} &=&  x_{\II}\sin\theta_\mathcal{B}\cos{\varphi_\mathcal{B}} + 		
	y_{\II}\sin\theta_\mathcal{B}\sin{\varphi_\mathcal{B}} + 
	z_{\II}\cos\theta_\mathcal{B} \nonumber
	\\
	y_\mathcal{B} &=& -x_{\II}(\cos\theta_\mathcal{B} \cos\varphi_\mathcal{B} \sin\alpha + 
	\sin\varphi_\mathcal{B} \cos\alpha)\nonumber \\ &&- 
	y_{\II}(\cos\theta_\mathcal{B} \sin\varphi_\mathcal{B} \sin\alpha - 
	\cos\varphi_\mathcal{B} \cos\alpha) +
	z_{\II}\sin\theta_\mathcal{B} \sin\alpha \nonumber 
	\\ 
	z_\mathcal{B} &=& -x_{\II}(\cos\theta_\mathcal{B}\cos{\varphi_\mathcal{B}}\cos\alpha - \sin\varphi_B\sin\alpha) \nonumber \\&& +    
	y_{\II}\sin\theta_\mathcal{B}\sin\alpha
	+z_{\II}\sin\theta_\mathcal{B} \cos\alpha.
\end{eqnarray}

The equation \ref{eq:elipsoide} can be, now, solved with respect to $\zeta$ giving
\begin{eqnarray}
	\zeta &=& R\left(1+E_1\mathcal{P}+E_2\mathcal{C}_2+E_3\mathcal{S}_2+E_4\mathcal{C}_1+E_5\mathcal{S}_1\right)
	\label{eq:zetaII}
\end{eqnarray}
where $\mathcal{P},\mathcal{C}_2,\mathcal{S}_2,\mathcal{C}_1,\mathcal{S}_1$ are the same functions of the angular coordinates of the surface point defined by Eqns. (\ref{eq:PCS})  
and the coefficients $E_i$ are parameters allowing to know the dynamic tide ellipsoid flattenings ${\mathcal E}_\rho,{\mathcal E}_z$ and the angles $\theta_\mathcal{B}$, $\varphi_\mathcal{B}$, and $\alpha$ which fix the orientation of the dynamic ellipsoidal bulge. They are
\begin{eqnarray}
\nonumber E_1 &=& \mathcal{A}-\mathcal{B}-3\mathcal{A}\cos^2{\theta_\mathcal{B}}+3\mathcal{B}\sin^2{\theta_\mathcal{B}}\sin^2{\alpha}, \\
	\nonumber E_2 &=& \Big(\mathcal{A}+\mathcal{B}-\mathcal{A}\cos^2{\theta_\mathcal{B}}-\mathcal{B}(1+\cos^2{\theta_\mathcal{B}})\sin^2{\alpha}\Big)\cos{2\varphi_\mathcal{B}}
	\\&&-\mathcal{B}\cos{\theta_\mathcal{B}}\sin{2\alpha}\sin{2\varphi_\mathcal{B}}, \nonumber \\
	E_3 &=& \Big(\mathcal{A}+\mathcal{B}-\mathcal{A}\cos^2{\theta_\mathcal{B}}-\mathcal{B}(1+\cos^2{\theta_\mathcal{B}})\sin^2{\alpha}\Big)\sin{2\varphi_\mathcal{B}}
	\\&&+\mathcal{B}\cos{\theta_\mathcal{B}}\sin{2\alpha}\cos{2\varphi_\mathcal{B}}, \nonumber\\
	\nonumber E_4 &=& \Big(\mathcal{A}+\mathcal{B}\sin^2{\alpha}\Big)\sin{2\theta_\mathcal{B}}\cos{\varphi_\mathcal{B}}+\mathcal{B}\sin{\theta_\mathcal{B}}\sin{2\alpha}\sin{\varphi_\mathcal{B}}, \\
 \nonumber E_5 &=& \Big(\mathcal{A}+\mathcal{B}\sin^2{\alpha}\Big)\sin{2\theta_\mathcal{B}}\sin{\varphi_\mathcal{B}}-\mathcal{B}\sin{\theta_\mathcal{B}}\sin{2\alpha}\cos{\varphi_\mathcal{B}},
	\label{eq:E_i}
\end{eqnarray}
with
\begin{equation}
	\mathcal{A} = \frac{1}{2}\Big(\frac{\mathcal{E}_\rho}{2}+\mathcal{E}_z\Big);\qquad
	\mathcal{B} = \frac{1}{2}\Big(\frac{\mathcal{E}_\rho}{2}-\mathcal{E}_z\Big).
\end{equation}

It is worth noting that both the parameters $f_i$ (given by Eqn. \ref{eq:fi}) and the parameters $E_i$ are linear functions of the flattenings $\epsilon_\rho,\epsilon_z$ and $\mathcal{E}_\rho,\mathcal{E}_z$, respectively. 

\section{The creep parametric equations}\label{sec:creep}

In the creep tide theory, the tidal deformation of the primary $\mathsf{m}$ is radial and given, in each direction $\varphi_s,\theta_s$ in the body by the Newtonian creep law
\begin{equation}\label{eq:creep}
\dot{\zeta}=\gamma(\zeta-\rho)
\end{equation}	
where $\zeta$ and $\rho$ are the corresponding radii vectors of the points on the surface of the dynamic and static tide and $\gamma$ is the relaxation factor (see Ferraz-Mello 2013). To obtain the explicit differential equations that define the dynamic figure of equilibrium, we have to introduce the expressions $\zeta$ and $\rho$ into Eqn.  \ref{eq:creep}. Since the body rotates around the polar axis of the frame $\mathcal{F}_{\rm II}$: $\dot{\theta_s}=0$ and $\dot{\varphi_s}=\Omega$. 
The time derivatives of the functions ${\mathcal{P}}, {\mathcal{C}_k}, {\mathcal{S}_k}$ (see Eqn. \ref{eq:PCS}) can be written as
\begin{equation}
\dot{\mathcal{P}} = 0; \quad
\dot{\mathcal{C}}_k = -k\Omega\mathcal{S}_k; \quad
\dot{\mathcal{S}}_k =   k\Omega\mathcal{C}_k \qquad\quad { (k=1,2)}.
\end{equation}		
Hence,
	\begin{equation} \label{eq:cumbersome}
\begin{array}{l} \displaystyle
	 [\dot{E}_1 - \gamma(f_1-E_1)] {\mathcal P} +  \\ 
	 
     [\dot{E}_2 - \gamma(f_2-E_2)+2\Omega E_3] {\mathcal C}_1   
	+[\dot{E}_3 - \gamma(f_3-E_3)-2\Omega E_2] {\mathcal S}_1 + \\
	
 	[\dot{E}_4 - \gamma(f_4-E_4) +\Omega E_5] {\mathcal C}_2
	 +[\dot{E}_5 - \gamma(f_5-E_5) -\Omega E_4] {\mathcal S}_2 = 0. \\
\end{array}
	\end{equation}

If we consider that the parameters $E_i$ cannot depend upon the coordinates of the generic direction 
where the calculation is done, the brackets in Eqn. \ref{eq:cumbersome} must independently vanish (Folonier et al. 2018) and 
we are left with the five first-order differential equations, the \textit{creep parametric equations}:
\begin{eqnarray}
	\nonumber \dot{E}_1 &=& \gamma(f_1-E_1);\\
	\nonumber \dot{E}_2 &=& \gamma(f_2-E_2)-2\Omega E_3;\\
	\dot{E}_3 &=& \gamma(f_3-E_3)+2\Omega E_2; \label{eq:dEi} \\
	\nonumber \dot{E}_4 &=& \gamma(f_4-E_4) -\Omega E_5;\\
	\nonumber \dot{E}_5 &=& \gamma(f_5-E_5) +\Omega E_4. 
\end{eqnarray}

Finding the solutions of Eq. (\ref{eq:dEi}) is equivalent to obtaining the orientation and shape of the instantaneous ellipsoid. The parameters $\mathcal{E}_\rho$, $\mathcal{E}_\rho$, $\varphi_\mathcal{B}$, $\theta_\mathcal{B}$ and $\alpha$ can be obtained by solving Eqn. (\ref{eq:E_i}), which can be easily done using an iterative procedure.   

The set of differential equations (\ref{eq:dEi}) is equivalent to the planar parametric system of equations found by Folonier et al (2018) when  $\theta=\theta_\mathcal{B}=\halfpi$ and $\alpha=0$. In this case, the polar parameters tend to the polar flattenings: 
$$E_1\rightarrow \mathcal{E}_z; \qquad f_1\rightarrow \epsilon_z,$$ 
while the variables $E_2$ and $E_3$ tend to the the equatorial variables:
$$E_2\rightarrow \frac{\mathcal{E}_\rho}{2}\cos{2\varphi_\mathcal{B}};\qquad
f_2\rightarrow \frac{\epsilon_\rho}{2}\cos{2\varphi};$$
$$E_3\rightarrow \frac{\mathcal{E}_\rho}{2} \sin{2\varphi_\mathcal{B}}; \qquad f_3\rightarrow \frac{\epsilon_\rho}{2}\sin{2\varphi}.$$ 
The resulting equations for $E_i$ (i=1,2,3) are the same equations given by Gomes et al. (2019) and Ferraz-Mello et al. (2020) for the coplanar case. 
In that case, the parameters $E_4$, $f_4$, $E_5$, $f_5$, and their time derivatives are all reduced to zero.

\section{The inertia tensor and the moment of quadrupole}\label{sec:I}

The inertia tensor $\mathsf{I}$ and the quadrupole moment $\mathsf{B}$ appear in the development of the tidal theories and will be defined in this section. Notwithstanding their conceptual differences, their comparison shows that they are mathematically related through
\begin{equation}
	\mathsf{I} = I_0(\mathbb{I}-\mathsf{B}), \label{eq:defB}
\end{equation}
(e.g., Ragazzo and Ruiz 2015, 2017, 2024; Bou\'e 2020; Folonier et al. 2022) where $\mathsf{I}$ is the moment of inertia of the body: \begin{equation}
	\mathsf{I}=\int_m \left(\begin{array}{ccc}
		(y^2+z^2)&-xy&-xz\\
		-xy&(z^2+x^2)&-yz\\ 
		-xz&-yz&(x^2+y^2)\\
	\end{array}\right)dm,	 
\end{equation}
$\mathbb{I}$ is the $3\times3$ identity matrix, $I_0=(I_{xx}+I_{yy}+I_{zz})/3=\textnormal{Tr}(\mathsf{I})/3$ 
and $\mathsf{B}$ is the nondimensional traceless moment of the quadrupole tensor.
It is not necessary to specify the frame used because the integral over the mass of the body is the same in all frames defined (which differ from each other by rotations only). 

Because of the tensor invariance, it is possible to calculate them in the frame \FB whose axes coincide with the principal axes of the dynamic tide ellipsoidal figure of equilibrium and then make the necessary rotations to obtain them in terms of the coordinates of the other frames. 

We thus have
\begin{equation}
	\mathsf{I}= \frac{1}{5}m \left(\begin{array}{ccc}
		(b_{\mathcal B}^2+c_{\mathcal B}^2)&0&0\\
		0&(c_{\mathcal B}^2+a_{\mathcal B}^2)&0\\ 
		0&0&(a_{\mathcal B}^2+b_{\mathcal B}^2)\\
	\end{array}\right),	 
\end{equation}
or, to first-order in the flattenings:
\begin{equation}
	\mathsf{I}= I_0 \left(\begin{array}{ccc}
		(1-\frac{1}{2}{\mathcal E}_\rho - \frac{1}{3}{\mathcal E}_z)&0&0\\
		0&(1+\frac{1}{2}{\mathcal E}_\rho - \frac{1}{3}{\mathcal E}_z)&0\\ 
		0&0&(1 + \frac{2}{3}{\mathcal E}_z)\\
	\end{array}\right),
\end{equation}
where  
\beq I_0 = \frac{1}{3} Tr({\mathsf{I}})= \frac{2}{5}mR^2 \endeq
and $m$, and $R$ are the mass and mean radius of the body.

\subsection{The moment of quadrupole in the equatorial frame}\label{sec:B}
From Eqn. (\ref{eq:defB}) we have, 
\beq \mathsf{B}=\mathbb I - I_0^{-1}\ \mathsf{I}, \endeq
that is,
\begin{equation}
	\mathsf{B}_\B= \left(\begin{array}{ccc}
		(\frac{1}{2}{\mathcal E}_\rho + \frac{1}{3}{\mathcal E}_z)&0&0\\
		0&(-\frac{1}{2}{\mathcal E}_\rho + \frac{1}{3}{\mathcal E}_z)&0\\ 
		0&0&(- \frac{2}{3}{\mathcal E}_z)\\
	\end{array}\right).
\end{equation}
To have the moment of quadrupole referred to the equatorial frame \FII, it is enough to use the corresponding rotation:
\beq \mathsf{B}_{\II}  = \mathsf{Q} \mathsf{B}_\B \mathsf{Q}^{-1} \endeq
(where the subscripts indicate the reference frame being used). Hence
\begin{equation}
		\mathsf{B}_{\II}= \left(\begin{array}{ccc}
	(\frac{E_1}{3} + E_2) & E_3 & E_4 \\
	E_3 & (\frac{E_1}{3} - E_2) & E_5\\ 
	E_4 & E_5 &  (- \frac{2E_1}{3}) \\
\end{array}\right)
\end{equation}
where  $E_i$ are the parameters of the dynamic tide defined in section \ref{sec:E_i}.

\section{The Dynamics. Force and torque}{\label{sec:Pot}}

The potential created by the homogeneous triaxial ellipsoid $\mathsf{m}$ at an external point ${\mathbf r}$, neglecting the harmonics of degree higher than 2, is:
\begin{equation}
	U({\mathbf r}) = -\frac{{\mathcal G}m}{r}+\frac{3{\mathcal G}}{2{r}^{5}}\Big({\mathbf r}\cdot \mathsf{I}{\mathbf r}-\frac{{r}^{2}}{3} \textnormal{Tr}(\mathsf{I})\Big),
	\label{eq:U}
\end{equation}
(see Beutler 2005) where $\mathsf{I}$ is the inertia tensor, $\textnormal{Tr}(\mathsf{I})$ its trace, {and $\mathcal{G}$ the gravitational constant}. The force acting on a companion of mass $M$ placed at ${\mathbf r}$ is
\begin{equation}
	{\mathbf F}({\mathbf r})= -M\nabla_{{\bf r}} U({\bf r}).
\end{equation}
The minus sign in this expression comes from the fact that $U$ is a potential, not a force function. Hence
\begin{equation}
	{\mathbf F}=-\frac{{\mathcal G}Mm}{r^3}{\mathbf r}
	-\frac{3\mathcal{G}M}{2r^5}\Big(2\mathsf{I}{\bf r}-\frac{5{\bf r}\cdot \mathsf{I}{\bf r}}{r^2}{\bf r}+\textnormal{Tr}(\mathsf{I}){\bf r}\Big)
	\label{eq:f disturbing}
\end{equation}
and the corresponding torque acting on the primary is ${\mathbf T}=-{\bf r}\times{\bf F}$, that is, 
\begin{equation}
	{\mathbf T}=\frac{3\mathcal{G}M}{r^5}{\bf r}\times\mathsf{I}{\bf r}.
	\label{eq:N torque}
\end{equation}

If we introduce the tensor $\mathsf{B}$ defined in sec.\ref{sec:B}, we have 
\begin{equation}
	U({\bf r}) = -\frac{{\mathcal G}m}{r} + \frac{3\mathcal{G}I_0}{2r^5}{\bf r}\cdot\mathsf{B}{\bf r}.
	\label{eq:dU(B)}
\end{equation}
Hence,
\begin{equation}
	{\mathbf F}=-\frac{\mathcal {G}Mm}{r^3}{\mathbf r} + \delta{\mathbf F}
 \end{equation}
where 
\begin{equation}
    \delta\mathbf{F} = \frac{3\mathcal{G}MI_0}{r^5}\Big(\mathsf{B}{\bf r}-\frac{5{\bf r}\cdot\mathsf{B}{\bf r}}{2r^2}{\bf r}\Big),
	\label{eq:F(B)}
\end{equation}
and 
\begin{equation}
	{\mathbf T}=-\frac{3\mathcal{G}MI_0}{r^5}{\bf r}\times\mathsf{B}{\bf r}.
	\label{eq:M(B)}
\end{equation}
We emphasize that all expressions in this section are independent of the chosen reference frame.

\section{The equations of motion}{\label{sec10}}

The equation of motion of M is
\begin{equation}
	\ddot{{\bf r}}_{[{\rm I}]} = -\frac{{\mathcal G}(m+M)}{r^3}{\mathbf r}_{[{\rm I}]}
	+\frac{{\delta\bf F}_{[{\rm I}]}}{\mu},
	\label{eq:force-newton1}
\end{equation}
where $\mu=Mm/(M+m)$ is the reduced mass of the two-body problem (see Ferraz-Mello et al., 2003) and  
\begin{equation}
    \delta\mathbf{F}_{[\rm I]} = \frac{3\mathcal{G}MI_0}{r^5}\Big(\mathsf{B_{[\rm  I]}}{\bf r_{[\rm I]}}-\frac{5{\bf r}\cdot\mathsf{B}{\bf r}}{2r^2}{\bf r_{[\rm I]}}\Big).
\end{equation}
The subscript $\I$ indicates that the given equation of motion is only valid in the reference frame $\mathcal{F}_{\rm I}$. No subscript was used in the term (${\bf r}\cdot\mathsf{B}{\bf r}$)   because it is a scalar quantity.

\section{Rotation equations}{\label{sec11}}

In the inertial axes reference frame $\mathcal{F}_{{\rm I}}$, the rotation equation is
\begin{equation}
	\frac{d}{dt}\Big(\mathsf{I}_{{\rm I}}{\bf \Omega}_{[{\rm I}]}\Big) = {\mathbf T}_{[{\rm I}]},
	\label{eq:torque-newton1}
\end{equation}
where $\mathsf{I}_{{\I}}$ and ${\bf \Omega}_{[{\rm I}]}$ are the inertia tensor and the angular velocity vector, both expressed in the astrocentric frame with inertial axes $\mathcal{F}_{{\rm I}}$. 

To have the equation in the reference frame $\mathcal{F}_{\rm II}$, we have to apply the transformation defined by Eqn. (\ref{eq:IItoI}) (and its inverse). Hence

\begin{equation}
	\frac{d}{dt}\big(\mathsf{I}_{\II}{\bf \Omega}_{[\rm II]}\big)+\mathsf{R}^ {\rm T}\dot{\mathsf{R}}\mathsf{I}_{\II}{\bf \Omega}_{[\rm II]} = {\mathbf T}_{[\rm II]}.
	\label{eq:dLII}
\end{equation}
If we linearize the equation $\dot{\bf \Omega}_{[{\rm II}]} = \frac{d}{dt}(\mathsf{R}^{-1}{\bf \Omega}_{[\rm I]})$, and neglect second-order contributions in the flattenings (or equivalently, in the parameters $E_i$), we obtain the following relations

\begin{eqnarray}
	\mathsf{I}^{-1}\dot{\mathsf{I}}{\bf \Omega} &=& -\dot{\mathsf{B}}{\bf \Omega}+ \mathcal{O}(\mathsf{B}^2),\nonumber\\
	\mathsf{I}^{-1}\mathsf{R}^ {\rm T}\dot{\mathsf{R}}\mathsf{I}{\bf \Omega} &=& \Big(\mathsf{R}^{\rm T}\dot{\mathsf{R}}+\mathsf{B}\mathsf{R}^ {\rm T}\dot{\mathsf{R}}-\mathsf{R}^ {\rm T}\dot{\mathsf{R}}\mathsf{B}\Big){\bf \Omega}+ \mathcal{O}(\mathsf{B}^2), \\
	\mathsf{I}^{-1}{\bf T} &=&  -\frac{3\mathcal{G}M}{r^5}{\bf r}\times\mathsf{B}{\bf r}+\mathcal{O}(\mathsf{B}^2), \nonumber
\end{eqnarray}
where we have omitted the subscript \II. All vectors and tensors in these final equations are assumed to be expressed in the equatorial frame $\mathcal{F}_{\rm II}$.

To the first order in the flattenings, the rotational equation is
\begin{equation}
	\dot{\bf \Omega}+\Big(\mathsf{R}^{\rm T}\dot{\mathsf{R}}+\mathsf{B}\mathsf{R}^ {\rm T}\dot{\mathsf{R}}-\mathsf{R}^ {\rm T}\dot{\mathsf{R}}\mathsf{B}\Big){\bf \Omega} = -\frac{3GM}{r^5}{\bf r}\times\mathsf{B}{\bf r}+\dot{\mathsf{B}}{\bf \Omega}.
	\label{eq:dot-omega}
\end{equation}
They can be expressed more compactly if we introduce the matrix $\mathsf{A}$ and the vector  $\dot{\mathsf{w}}$:
\begin{equation}
	\mathsf{A} =  
	\left(\begin{array}{ccc}
		(1+E_1+E_2)\sin{J}+E_5\cos{J}  &  E_3 &  0 \\
		E_3\sin{J}-E_4\cos{J}  &  1+E_1-E_2 & 0 \\ 
		2E_4\sin{J}  &  2E_5 &  1 \end{array} \right);\qquad \dot{\mathsf{w}}= \left(\begin{array}{c}
		\Omega\dot{\psi} \\
		-\Omega\dot{J} \\
		\dot{\Omega} \end{array} \right). 
\end{equation}
Hence, from Eqn. (\ref{eq:dot-omega}),
\begin{equation}	\label{eq:detA}
	\mathsf{A}\dot{\mathsf{w}} = -\frac{3GM}{r^5}{\bf r}\times\mathsf{B}{\bf r}+\dot{\mathsf{B}}{\bf \Omega}.
\end{equation}

The first-order differential equations (\ref{eq:force-newton1}), (\ref{eq:detA}), and (\ref{eq:dEi}) compose the complete set of differential equations that describe the motion of the companion and the rotation and orientation of the primary, respectively.

The matrix $\mathsf{A}$ is trivially decomposable, and Eqn. (\ref{eq:detA}) can be divided into two parts:

\begin{displaymath}
	\Omega\left(\begin{array}{cc}
		(1+E_1+E_2)\sin{J}+E_5\cos{J}  &  E_3 \\
		E_3\sin{J}-E_4\cos{J}  &  1+E_1-E_2 \\ 
	\end{array} \right)\left(\begin{array}{c}
		\dot{\psi} \\
		-\dot{J} \\
	\end{array} \right)  = 
\end{displaymath}
\begin{equation}
	\left[-\frac{3GM}{r^5}{\bf r}\times\mathsf{B}{\bf r}+
	\dot{\mathsf{B}}{\bf \Omega}\right]_{(1,2)}
\end{equation}
and
\begin{equation}
	\dot\Omega=\left[-\frac{3GM}{r^5}{\bf r}\times\mathsf{B}{\bf r}+
	\dot{\mathsf{B}}{\bf \Omega}\right]_{(3)} -2\Omega E_4\dot\psi\sin J + 2\Omega E_5\dot{J},
	\label{eq:rot}
\end{equation}
where the subscripts (1,2) and (3) mean the two first and the third vector components, respectively.

\section{Application: The dynamic tide of CoRoT-3b}\label{Part4}

CoRoT-3b is a massive extrasolar companion, a brown dwarf star orbiting an F-type star. It is one of the various dwarf companions discovered by the space mission CoRoT (Deleuil et al. 2008). The orbit is inclined with respect to the stellar equator, and it has been thoroughly observed from many ground telescopes. As a result, the main orbital and rotational elements are known (Table \ref{tab:Corot}). The stellar rotation is faster than the orbital period, indicating that the synchronization has not yet been reached. The value 3.65 d obtained by Tadeu dos Santos (comm. pers.) from a reanalysis of the Rossiter-McLaughlin effect in the observations done by Triaud et al. (2009), is reinforced by the existence of a peak at about 3.6 days in the power spectrum of the photometric signal.

\begin{table}
	\caption{The extrasolar system CoRoT-3}\label{tab:Corot}
	\begin{tabular}{lcc}
		\hline\\
		& Star  & CoRoT-3b  \\
		\hline
		Mass &	$1.41 \pm 0.08^{(1)} \ M_\odot$   &   $21.66 \pm 1^{(2)}  \ M_{\rm Jup}$  \\
		Radius & $1.44 \pm 0.08^{(1)} \ R_\odot$   &    $ 1.19 \pm 0.1^{(2)}\ R_{\rm Jup}$  \\
		Rotation Period$^{(3)}$ & $ 3.65 d $ &\\
		Age$^{(4)}$ & $1.6-2.8$ Gyr  \\
		Relaxation factor $\gamma^{(5)}$ \qquad \qquad & $40-70\ s^{-1}$ \\
		\hline
		&  \multicolumn{2}{c}{Orbital elements}\\
		\hline
		Semi-major axis$^{(2)}$&    \multicolumn{2}{c}{$0.05783 \pm 0.00085$ AU}\\
		Stellar obliquity$^{(6)}$ &  \multicolumn{2}{c}{$37.6^{+10}_{-22.3}$ deg} \\
		Eccentricity$^{(6)}$ &\multicolumn{2}{c} {$0.003 - 0.023$} \\
		Orbital Period$^{(7)}$ &\multicolumn{2}{c} {4.2567994 day} \\
		\hline
		\multicolumn{3}{l} {(1) Tsantaki et al. 2014; (2)Johns et al. 2018; 
			(3) (photometric) Tadeu dos Santos (priv. comm.); }\\
		\multicolumn{3}{l} 	{(4) Deleuil et al. 2008; (5) Ferraz-Mello, 2016; (6) Triaud et al. 2009;
			(7) Encyclopaedia of}\\
		\multicolumn{3}{l} {Exoplanetary Systems. http://exoplanet.eu} \\
	\end{tabular}
\end{table}

Simulations of the evolution of the rotational period of CoRoT-3 and
of the orbital period of the companion CoRoT-3b with a coplanar
tide model (Ferraz-Mello, 2016) showed that to reach the
observed stellar rotation in a time corresponding to the age of the system
(1.6 $-$ 2.8 Gyr), the relaxation factor $\gamma_{\rm st}$ must be between 40 and 70 s$^{-1}$.

\section{The dynamic equilibrium tide.}

In this section, we discuss the variation of the parameters of CoRoT-3b and its host star using the full 3D equations developed in this paper and the physical parameters given in Table \ref{tab:Corot}. The unknown parameters are fixed arbitrarily. In particular, the rotations of CoRoT-3b are initially assumed to be synchronous with its orbital motion, and the companion's obliquity is set close to the stellar obliquity. 

\begin{figure}[h]
	\begin{center}
		\includegraphics[width=10cm]{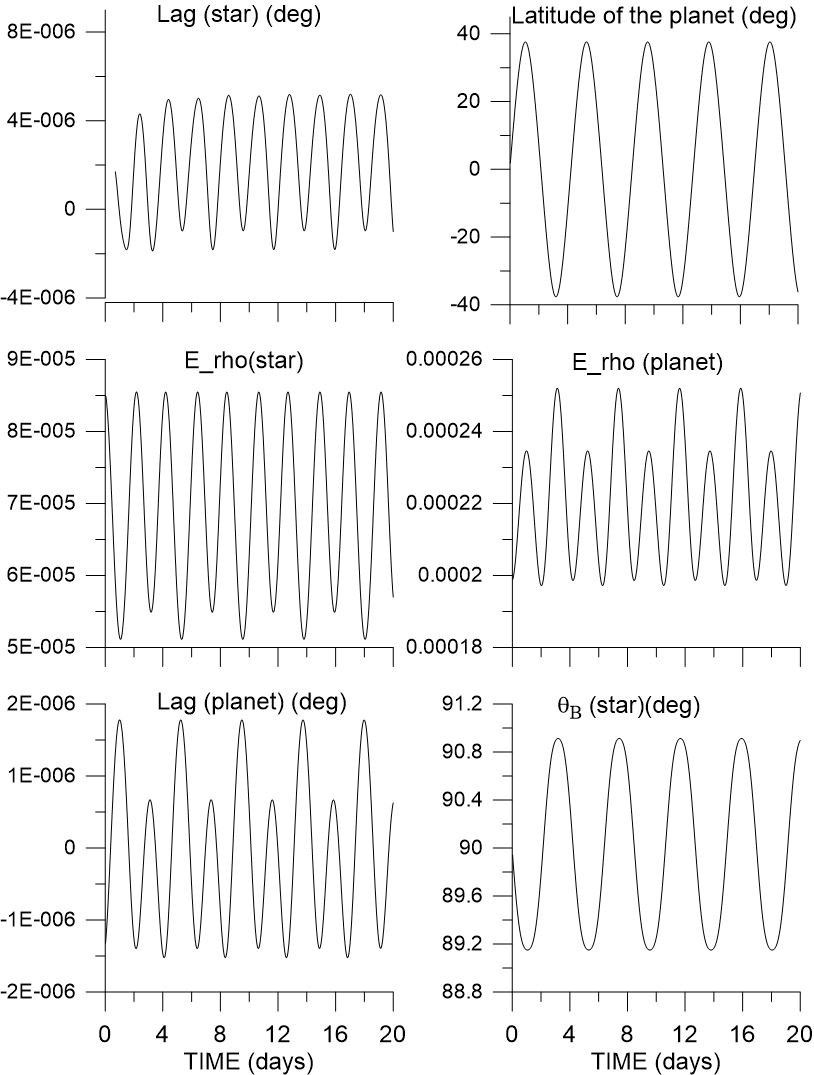}
		\caption{Short-term evolution of some parameters of CoRoT-3b and its host star: The tidal lags, the relative height of the tidal bulges ({adimensional}), and the co-latitude of the star tidal bulge. The latitude of the planetary companion with respect to the star's equator is also given for reference.}
		\label{fig:short72}
	\end{center}
\end{figure}

The main problem in the integration of the differential equations is the stiff character of the parametric equations. It is easy to show that the solutions obtained by finite increments (as is done in numerical integrations) are divergent, unless the step is kept very small -- less than $1/\gamma$. The adopted value of $\gamma_{\rm st}$ (70 s$^{-1}$) implies steps of hundredths of one second. This introduces a serious limitation to the numerical experiments and we had to limit the timespan of the integrations. Some tens of days are enough to see how the tides deform the figure of the bodies during some orbits. The results are shown in Fig. \ref{fig:short72}\footnote{In the figures of these sections, the standard label \textit{planet} is used to designate the planetary companion even if it is a brown dwarf.}.

The short-term evolution of the figure of the bodies (the so-called dynamical tide) is marked by the main characteristics of the system: the periods' almost double synchronization, the high stellar obliquity, and the low eccentricity. In the upper part of Fig. \ref{fig:short72}, the lag of the star is shown. It is mostly positive ($\varphi_B>\varphi$). The bulge is driven forward by the star's rotation, thus creating a torque contrary to the rotation direction. The star rotation is being braked by the tide. But this is a process that must be at work since the beginning of the life of the star and the rotation is approaching synchronization (see Ferraz-Mello, 2016). However, it is not far from synchronization and the lag is oscillating between positive and negative values, as normally seen in synchronously rotating stars. The lag is dominated by an oscillation whose period is half the orbital period. This is characteristic of the inclination perturbation effects (not only tidal ones) that always appear with twice the latitude as the argument. The imperfect periodicity (the successive oscillations are not exactly equal) must be due to the eccentricity component of the tide, which is not zero, has a period equal to the orbital period, and affects unequally successive oscillations. For the sake of comparison, the latitude of the planetary companion is shown on top of the figure.

In the middle row of the figure, we show the relative height (high tide $-$ low tide) of the tide in both the star and the planetary companion. They show oscillation characteristics similar to the lag. However, in planetary lag, one of the oscillations has a lower amplitude. This must be due to the particular values adopted for the angular initial conditions that were fixed arbitrarily.

\begin{figure}[h]
	\begin{center}
		\includegraphics[width=10cm]{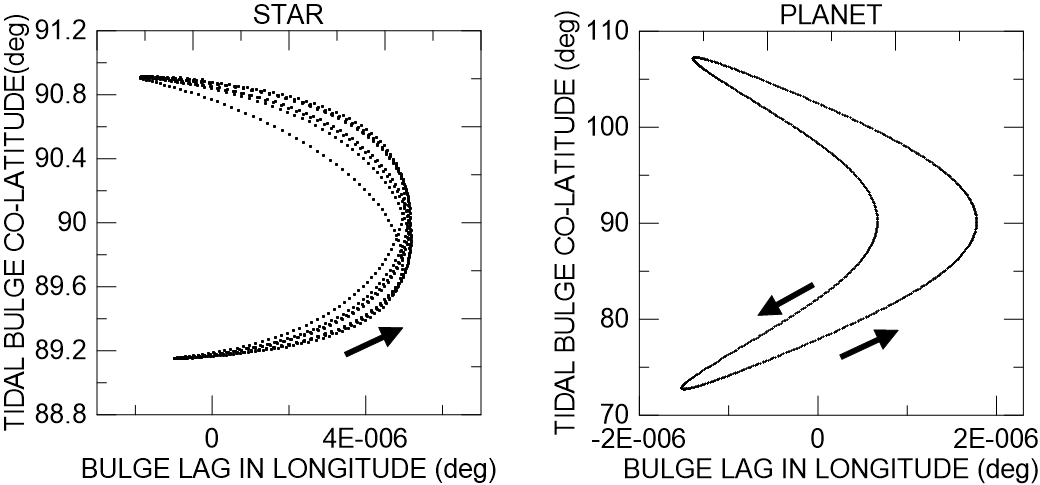}
		\caption{Motion of the tidal bulges. Left: the star. Right: CoRoT-3b. The coordinates are the tidal lag $\varphi_\B-\varphi$ and the co-latitude $\theta_\B$. }
		\label{fig:KG72}
	\end{center}
\end{figure}

The left plot in the last row of the figure shows the lag of the tidal bulge of CoRoT-3b (the ``planet"). 
The motion of the tidal bulge is better seen in a figure showing the two components together (Fig. \ref{fig:KG72} right). The lag in longitude is very small and the bulge is rather moving close to the sub-stellar meridian, from one hemisphere to another.  The same also happens with the star's tidal bulge (Fig. \ref{fig:KG72} left).

The general solution of the parametric equations is given by one transient term proportional to $e^{-\gamma t}$ and a series of forced terms. Since $\gamma$ is very small,
this transient is damped in a short time and only the forced terms matter in evolution studies. In analytical theories, we discard the transient, but in numerical simulations, there is no way to separate it from the forced terms, other than waiting for them to be naturally damped.

One variable not shown in the figures is the alignment angle $\alpha$. All numerical experiments have shown that it is not significantly different from zero. This means that in all our examples, the angular rotation vector lies in the plane defined by the major and minor axes of the triaxial ellipsoidal body figure (the vertex of the bulge and the pole of the figure) as in the case of the static tide as shown by Folonier et al. (2022). 
We have ruled out the possibility of it being an artifact of the iterative procedure used to solve Eqns. (\ref{eq:E_i}). The result is persistent and appears in all simulations performed. We verify directly from the output of the integrations that
\begin{equation}
	\Xi=\frac{E_3}{E_2}(\frac{E_4}{E_5}-\frac{E_5}{E_4})-2 \simeq 0,
	\label{eq:Xi}
\end{equation}
and the analysis of this equation shows that it may only happen if $\alpha \simeq 0$.
We note that
a similar constraint exists with the static tide parameters $f_i$:
\begin{equation}
\frac{f_3}{f_2}(\frac{f_4}{f_5}-\frac{f_5}{f_4})-2=0.
\end{equation}

If we force the initial conditions of the integration to be such that $\Xi \ne 0$, the solutions are such that, after a few steps, we have $\Xi\simeq 0$. 
\begin{figure}[t!]
	\begin{center}
		\includegraphics[width=10cm]{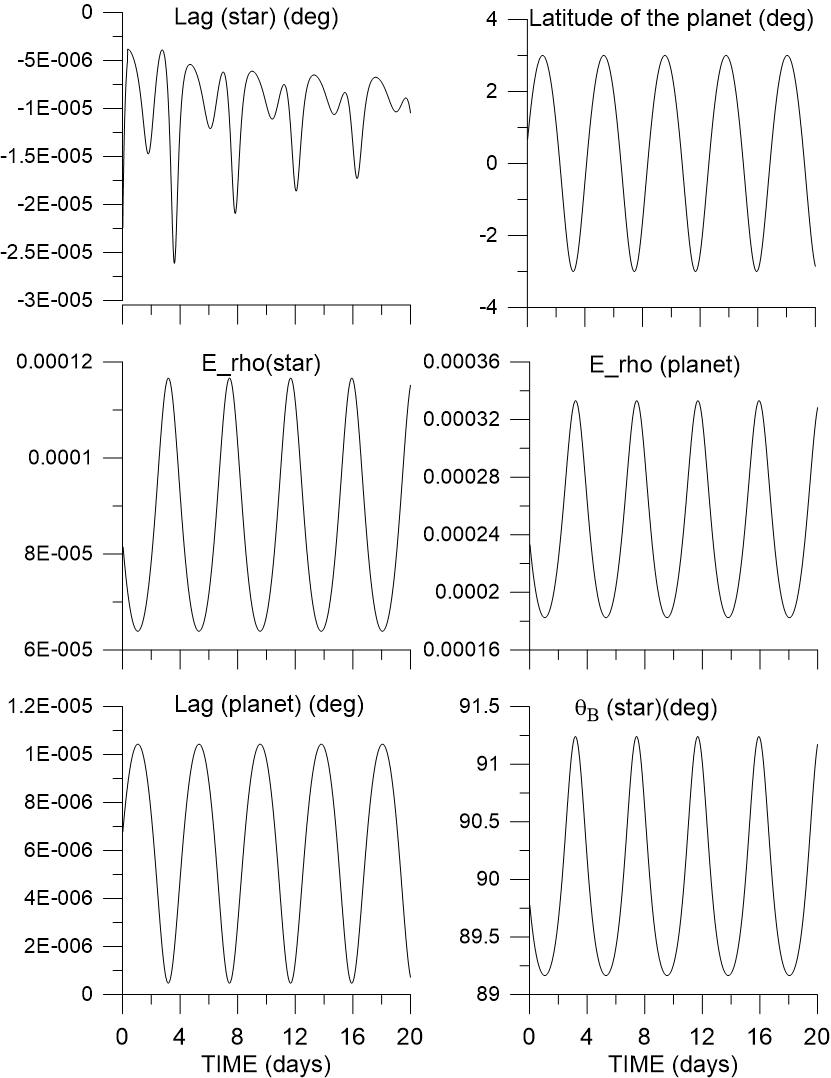}
		\caption{Short-term evolution of the parameters of a contrasting clone system with a super-synchronous planetary companion in an elliptic orbit around a slow-rotating star and low stellar obliquity. Same description as Fig. \ref{fig:short72}.}
		\label{fig:short77}
	\end{center}
\end{figure}

\section{A contrasting clone example}

To better assess the results shown in the previous figures, we present the results obtained for a contrasting clone system. In this system, we keep the basic parameters of CoRoT-3 but assume a supersynchronous planetary companion ($P_{\rm rot}= 0.8 P_{\rm orb}$)
in orbit around a slow-rotating star ($P_{\rm rot}=14.6$ d) with eccentricity $e=0.1$ and low stellar obliquity ($3^\circ$) (Fig. \ref{fig:short77}).

\begin{figure}[h]
	\begin{center}
		\includegraphics[width=10cm]{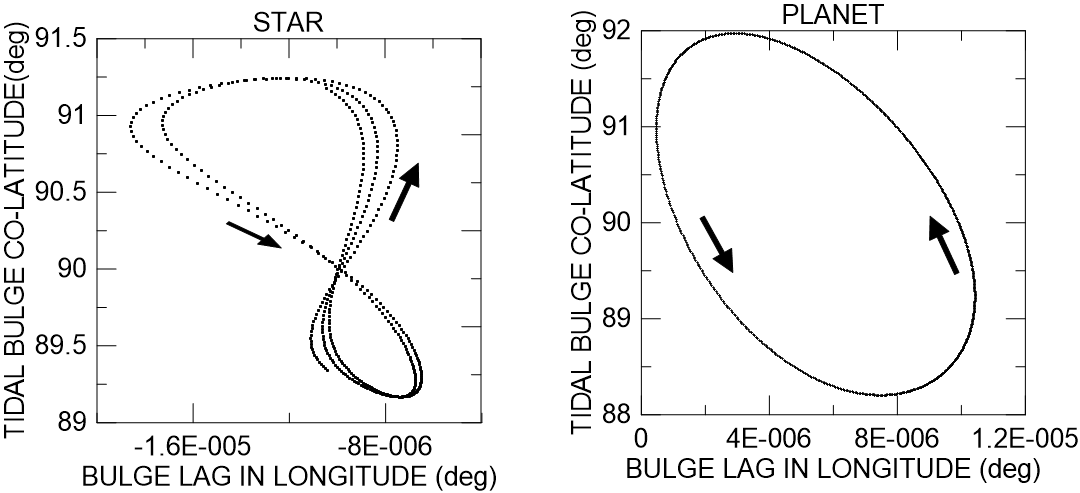}
		\caption{Motion of the tidal bulges in the clone system. Left: the star. Right: the planetary companion. }
		\label{fig:KG77}
	\end{center}
\end{figure}

 The lag of the star is now always negative. The bulge is driven backward, creating a torque in the rotation direction. Angular momentum is being transferred from the orbital motion to the star's rotation, accelerating it. In contrast, the planet's lag is always positive, indicating that the planet's rotation is being braked. All parameters shown in the figure oscillate with a period equal to the orbital period. This is a consequence of the dominance of the distance star-planet variation and the orbit's eccentricity. The transitory behavior seen in some figures is due to the arbitrary choice of the initial conditions. 

As in the actual CoRoT-3 system, the motion of the tidal bulges in the clone system is shown in Fig. \ref{fig:KG77}. The lag in longitude of CoRoT-3b is always positive, and the substellar meridian is on the left border of the figure. In the case of the star, the lag is negative.  

  \section{Orbital elements, rotations, and obliquities}\label{sec:longperiod}

\begin{figure}[ht]
	\begin{center}
		\includegraphics[width=10cm]{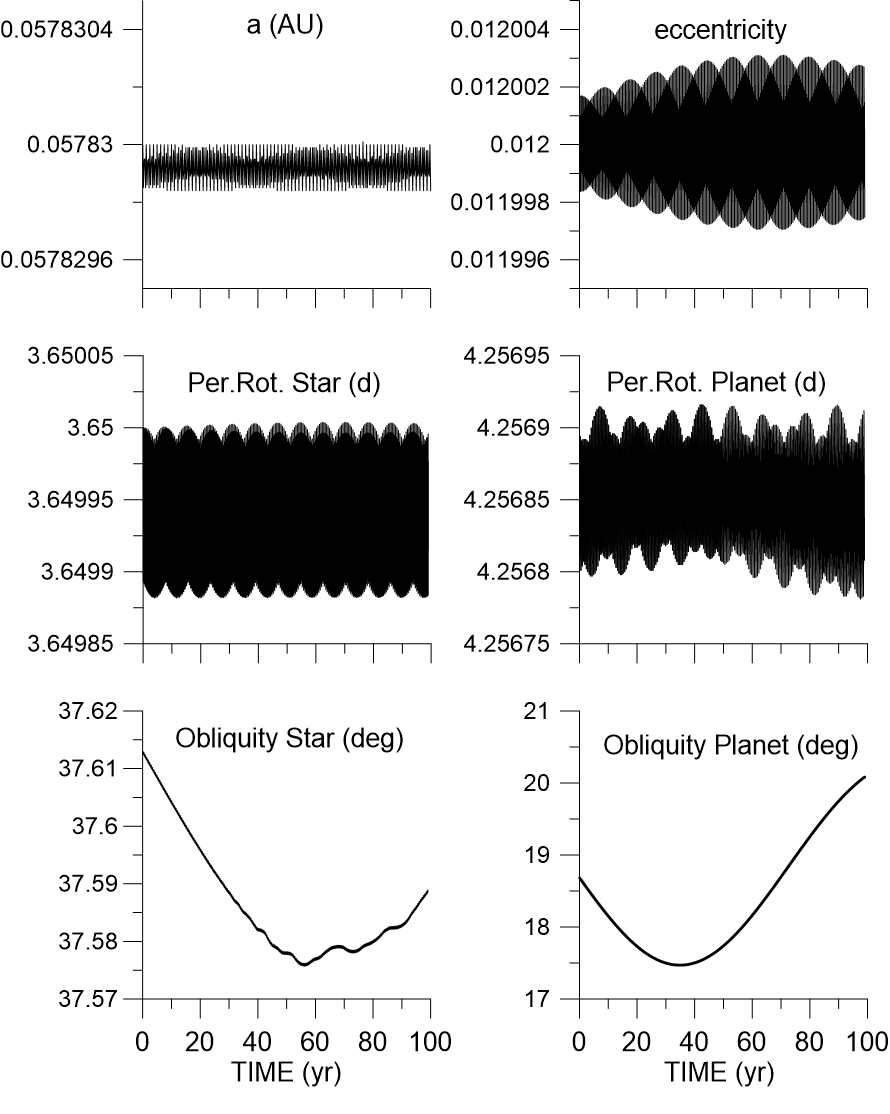}
		\caption{{Variation of the semi-major axis, eccentricity, rotation periods, and obliquities in the system CoRoT 3. }}
		\label{fig:elem}
	\end{center}
\end{figure}

It is worth calculating the variations of the main orbital, rotational, and geometric elements. However, the star and the brown dwarf companion have masses strongly concentrated in their central regions, while the theory established here refers to homogeneous bodies. Given the nonavailability of a fully layered model, we may introduce the actual Moment of Inertia factors in the tidal forces to cope with this difficulty. We adopted, for the star, the Moment of Inertia factor of the Sun: $I_0/mR^2=0.07$, and for the brown dwarf CoRoT-3b, $I_0/mR^2=0.1$, which is consistent with the Love number $k_2=0.387$ (Becker et al. 2018). 
In the short time considered, the orbital elements and the rotations show only large periodic variations. They are given in the top plots of Fig. \ref{fig:elem}.
The obliquities ($\rm \beta$) themselves do not appear in the set of variables considered by the theory. The angles that appear there are the inclinations of the orbit ($I$) and of the equators ($J$) with respect to the fundamental plane of the astrocentric frame \FI. The obliquities are to be obtained from the geometry of the system (see Fig. \ref{fig:H-6-tri} in the Appendix). The polar extension of the cosine law of spherical trigonometry applied to the spherical triangle $\mathsf{N}\mathsf{N}'\mathsf{N}_e$ gives
\beq
\cos\beta=  \cos J \cos I+\sin J \sin I \cos\Omega_{\rm  I}.
\endeq
The variation in obliquities is shown in the bottom panels of Fig. \ref{fig:elem}.

\section{Conclusion}

The creep tide theory was extended to exoplanetary systems with significant stellar and planetary obliquities. The new theory is complete and allows us to obtain the stellar and planetary hydrodynamic figures of equilibrium of the two bodies, their rotational dynamics, and the orbital perturbations. As in the old coplanar theory, the full 3D theory considers only first-order effects keeping one body unaltered (the companion) while calculating the tidal effects of the other (the primary). However, no hierarchy is imposed, and the theory that gives the tides on one body is the same that does it on the other. We need just change who is the primary. In the application, both cases were considered. 

The theory was applied to the system formed by the star CoRoT-3 and its close-in planetary companion, the brown dwarf CoRoT-3b. The use of this system as an example was motivated by several reasons: (i) the large mass of the brown dwarf (21.66 M$_{\rm Jup}$) and the proximity of the bodies ($<0.06$ AU) enhance the tidal effects on the star; (ii) the CoRoT exoplanetary systems have been subject to prolonged observational follow-ups that have allowed many of their physical parameters to be well determined; (iii) the stellar obliquity has been measured (Triaud et al. 2009) and is relatively high. In addition, the star CoRoT-3 has been the subject of studies allowing us to have good estimates of its age and relaxation. 

The results of the simulations performed were thoroughly discussed in Sections \ref{Part4} -- \ref{sec:longperiod}. We mention the heights of the tidal protrusion (high tide minus low tide) and the paths of the tidal bulges on the surfaces of the star and planet. We also emphasize those results that may be determinant for future investigations using the analytic solution of the parametric equations such as the almost vanishing alignment angle $\alpha$. It has a geometric interpretation: if $\alpha=0$, the spin axis of the body may lie in one of the symmetry planes of the dynamical equilibrium figure,
between the pole of the figure (minor vertex of the dynamic equilibrium ellipsoid) and the tidal bulge (major vertex of it).
This situation is equal to what happens in the static equilibrium tide, but we have not found a physical reason for it. The other parameters did not show significant variations in the short period of the simulations allowed by the stiffness of the equations. 

This study of the deformations of the bodies is the first step toward the study of the long-term evolution of the system. The next steps are to average the right-hand sides of the parametric equations and construct analytical solutions that allow us to reduce the number of degrees of freedom of the system of differential equations and overcome its stiffness. These steps will be presented in a forthcoming article.

\appendix

\section{Appendix: Reference frames and coordinate transformations}
The creep equation, the rotation equations, and the equations of motion are
simple. However, they can quickly become cumbersome due to the different
systems of reference used to express the main quantities involved. In this appendix, we describe these frames of reference, the notation used for each of them, and the rotation matrices that allow the respective transformations of coordinates.

\subsection{Astrocentric reference frame with inertial axes \FI}\label{sec:FI}
\begin{figure}[th]
	\begin{center}
			\includegraphics[width=10cm]{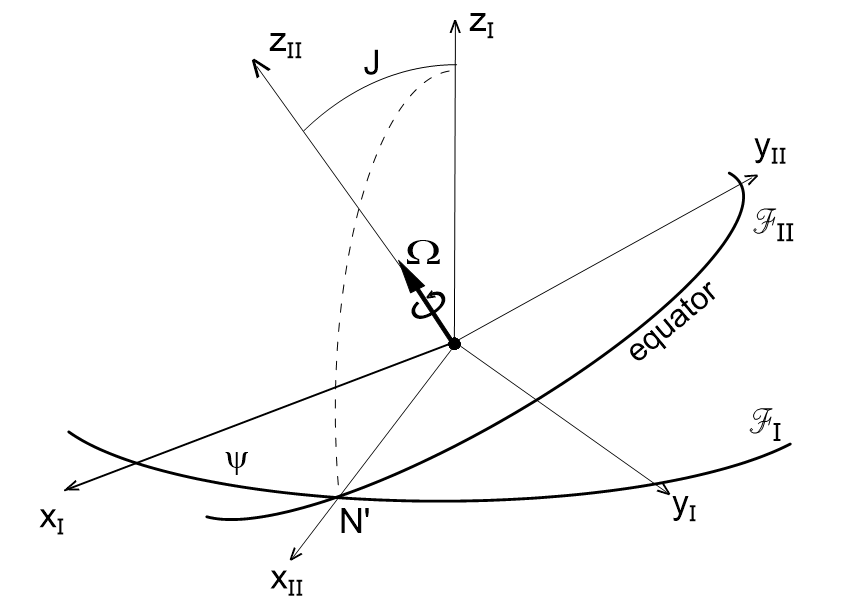}
		\caption{Astrocentric ($\mathcal{F}_{\rm I}$) and equatorial ($\mathcal{F}_{\rm II}$) reference frames.}
		\label{fig:H-3}
	\end{center}
\end{figure}

The reference frame \FI is centered on the primary, and their axes have inertial directions. It is sometimes dubbed inertial, but only the axes' directions are inertial. The origin follows the motion of the body (the system is astrocentric). In this reference frame, we define the angles $\Psi$  and $J$ as the longitude of
the ascending node of the equator $\mathsf{N}'$  and the co-latitude of the rotation pole defined by the angular
velocity vector $\Omega$, respectively. (Fig.\ref{fig:H-3}).

\subsection{Equatorial reference frame \FII}\label{sec:FII}
The equatorial reference frame \FII is a reference frame centered on the
primary, where the $z_{\rm  II}$-axis and the $x_{\rm  II}$-axis point in the direction of the angular velocity vector and the direction of the ascending node
of the equator $\mathsf{N}'$, respectively
(Fig. \ref{fig:H-3}).

\begin{figure}[t]
	\begin{center}
				\includegraphics[width=10cm]{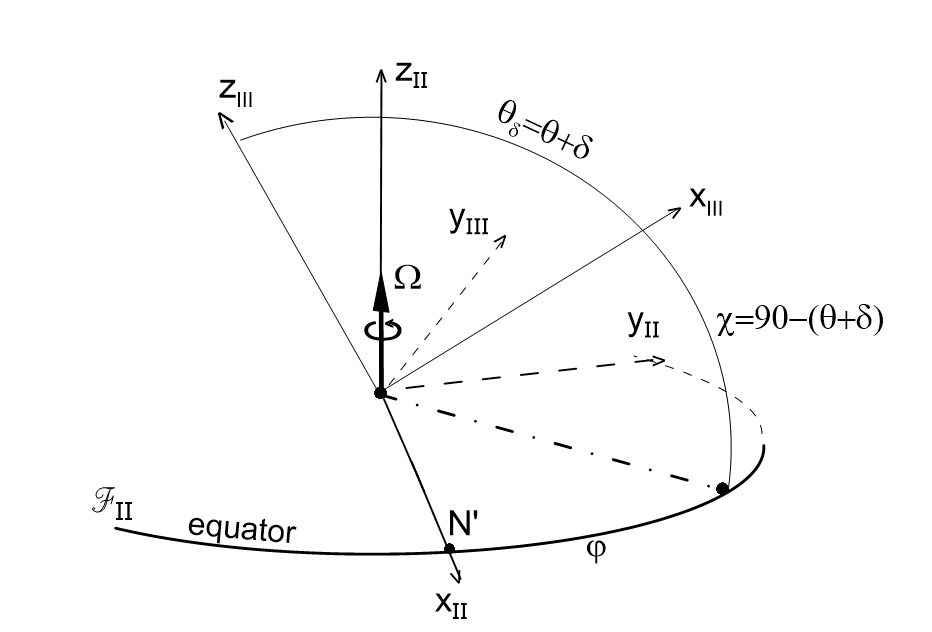}
		\caption{Equatorial reference frame \FII and the static equilibrium reference frame \FIII fixed to the
			symmetry axes of the static tide equilibrium ellipsoid. The $z_{\rm II}$-axis and the $x_{\rm II}$-axis are pointing in the direction of the rotation pole and of the ascending node
			of the equator $\mathsf{N}'$, respectively.}
		\label{fig:H-4}
	\end{center}
\end{figure}

In this reference frame, the longitude and co-latitude of the companion are, respectively, $\varphi$ and $\theta$ (see Fig. \ref{fig:H-4})

The rotation matrix leading from $\mathcal{F}_{\rm I}$ to $\mathcal{F}_{\rm II}$ is given as
\beq
\mathsf{R} = \mathsf{R}_3(\psi)\mathsf{R}_1(J),
\endeq
where $\mathsf{R}_j$ $(j=1,2,3)$ are the classical rotation matrices around the first, second, and third axes, respectively\footnote{
	$\mathsf{R}_1(\alpha)=\left( \begin{array}{ccc} 1&0&0\\0&\cos\alpha&-\sin\alpha\\0&\sin\alpha&\cos\alpha \end{array} \right)$;\quad
	$\mathsf{R}_2(\alpha)=\left( \begin{array}{ccc} \cos\alpha&0&\sin\alpha\\0&1&0\\-\sin\alpha&0&\cos\alpha \end{array} \right)$;\quad
	$\mathsf{R}_3(\alpha)=\left( \begin{array}{ccc} \cos\alpha&-\sin\alpha&0\\\sin\alpha&\cos\alpha&0\\0&0&1 \end{array} \right)$.}.
The transformation of the components of one vector $\mathbf{v}$ in the frame \FII to the components in the frame \FI is
\beq
\mathbf{v}_{[\rm I]}= \mathsf{R}\mathbf{v}_{[\rm II]}	\label{eq:IItoI}
\endeq
where the subscripts indicate the considered reference frame. 

\subsection{Static equilibrium reference frame \FIII}\label{sec:FIII}
The static equilibrium reference frame \FIII is centered on the primary and its axes are along the major and minor axes of the static (inviscid) equilibrium ellipsoid. 
The $x_{\rm III}$-axis lies along the longest axis.
The $z_{\rm III}$-axis lies along the shortest axis. 
The angular velocity vector and the radius vector of the companion
are in the $x_{\rm III} z_{\rm III}$-plane (Folonier et al. 2022). See Fig. \ref{fig:H-4} (see also Fig. \ref{fig:H-2} in Sec. \ref{sec:static}).
The longitude and the co-latitude of the $x_{\rm  III}$-axis in the reference frame \FII are, respectively, $\varphi$ and $\theta_\delta=\theta+\delta$.  The angle $\theta$ is the co-latitude of the companion $\mathsf{M}$ in the reference frame \FII and $\delta$ is the angle between the radius vector of $\mathsf{M}$ and the major axis of the static ellipsoid.

\begin{figure}[t]
	\begin{center}
				\includegraphics[width=11cm]{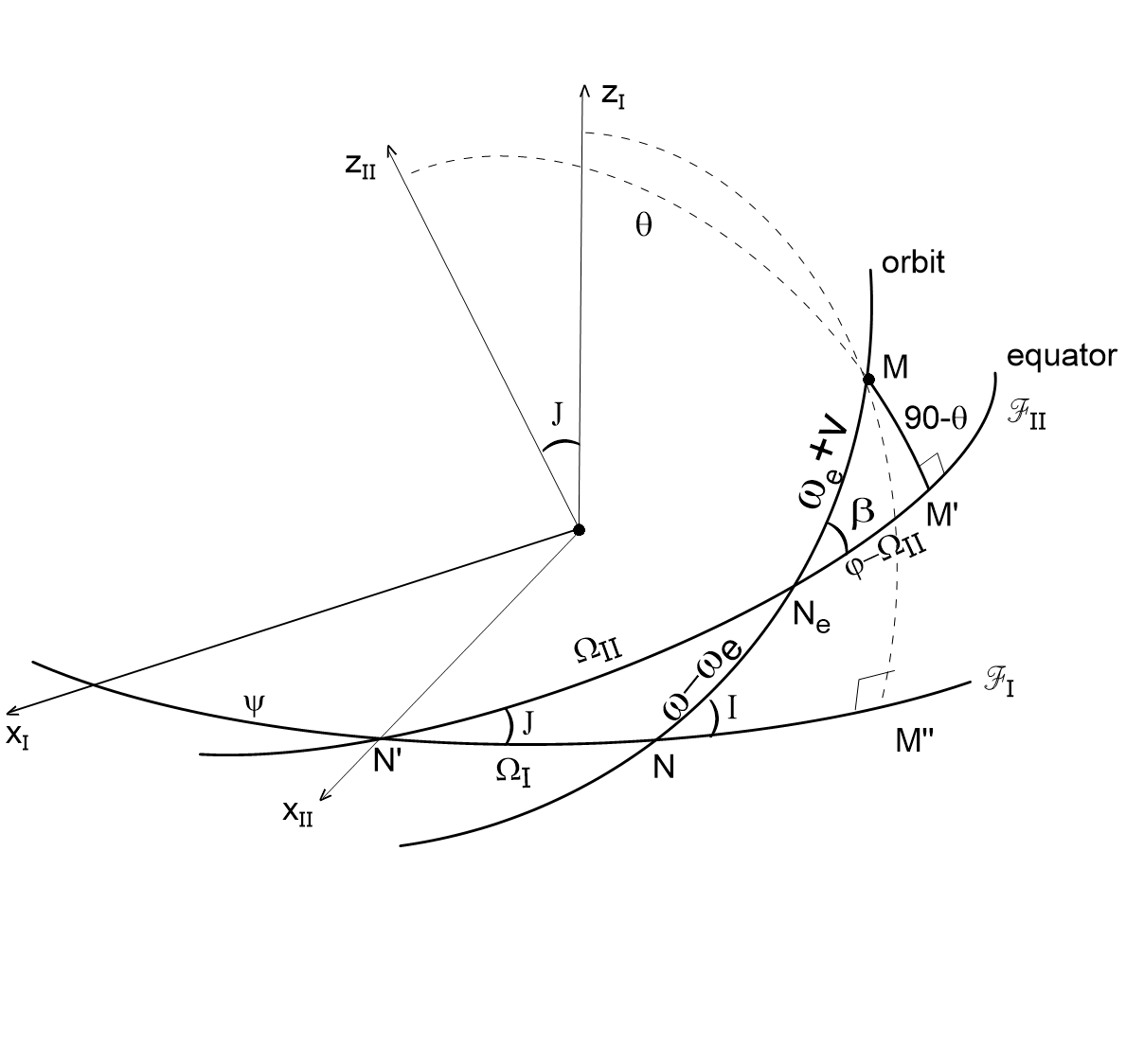}
		\caption{The main planes: fundamental plane of the astrocentric frame \FI, equator, and orbital plane.
			The angle $\beta$ is the obliquity of the primary.}
		\label{fig:H-6-tri}
	\end{center}
\end{figure}

The rotation matrix leading from $\mathcal{F}_{\rm II}$ to $\mathcal{F}_{\rm III}$ is written as
\begin{equation}
\mathsf{P} = \mathsf{R}_3(\varphi)\mathsf{R}_2(\theta_\delta-\halfpi),
\end{equation}
while the transformation of the components of one vector  in the frame \FIII to those in the frame \FII is
\begin{equation}
\mathbf{v}_{[\rm II]}= \mathsf{P}\mathbf{v}_{[\rm III]}.\label{eq:IItoIII}
\end{equation}

It is important to note that, in general, the axes of this ellipsoid, defined by the surface $\rho(\varphi_s,\theta_s)$ (see Sec. \ref{sec:static}), do not coincide with the principal axes of inertia of the primary, defined by the surface $\zeta(\varphi_s,\theta_s)$ (see Sec. \ref{sec:dynamic}). 

These transformations are summarized in Fig. \ref{fig:H-6-tri}. We avoided identifying all angles and arcs that appear in this figure to prevent a noxious overload. 

\subsection{The body reference frame \FB}\label{sec:FB}
The body reference frame \FB is centered on the primary. The axes $x_{\mathcal{B}}$ and $z_{\mathcal{B}}$ are pointing in the directions of the major and minor axes of the dynamic ellipsoidal figure of equilibrium of the primary, respectively, on the same side as the corresponding axes in frame $\mathcal{F}_{\rm III}$. The angles $\varphi_{\mathcal{B}}$ and $\theta_{\mathcal{B}}$ are the longitude and the co-latitude of the direction $x_{\mathcal{B}}$, given in the reference frame $\mathcal{F}_{\rm II}$, respectively. 

\begin{figure}[t]
	\begin{center}
				\includegraphics[width=10cm]{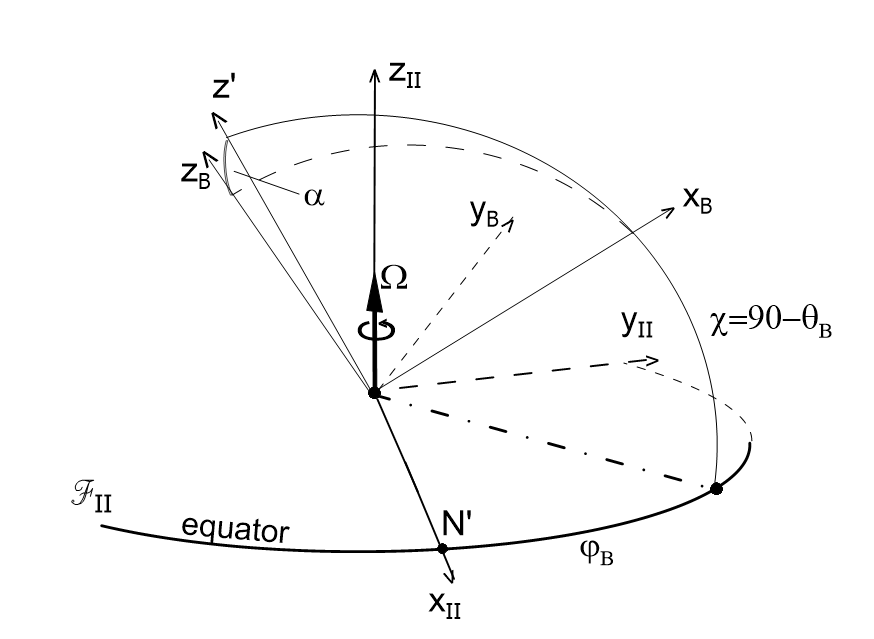}
		\caption{Equatorial reference frame \FII and the dynamic equilibrium reference frame \FB fixed to the
			symmetry axes of the dynamic tide equilibrium ellipsoid. The $z_\mathcal{B}$-axis and the $x_\mathcal{B}$-axis lie along the minor and major axis of the 
			ellipsoid, respectively.}
		\label{fig:H-B}
	\end{center}
\end{figure}

This frame is very similar to the frame \FIII but, at variance with that frame, the $z_{\mathcal{B}}$-axis is not in the meridian plane defined by $x_{\mathcal{B}}$ and the rotation pole (Fig. \ref{fig:H-B}). However, one rotation around the $x_{\mathcal{B}}$-axis can drive the $z_{\mathcal{B}}$-axis to the meridian plane and to a system similar to $\mathcal{F}_{\rm III}$. (The $y_{\mathcal{B}}$-axis also is driven to a new position, not shown in the figure.) 

\begin{figure}[h!]
	\begin{center}
				\includegraphics[width=10cm]{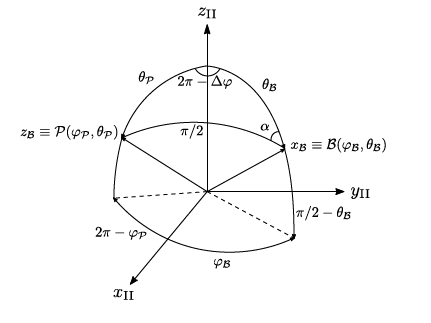}
		\caption{Spherical triangle used to calculate the angle $\alpha$. The points $\mathcal{B}$ and $\mathcal{P}$ are the directions of the $x_\mathcal{B}$-axis and $z_\mathcal{B}$-axis. Their longitude and co-latitude are, respectively, ($\varphi_\mathcal{B}$,$\theta_\mathcal{B}$), and ($\varphi_\mathcal{P}$,$\theta_\mathcal{P}$), with $\Delta\varphi= \varphi_\mathcal{P} - \varphi_\mathcal{B}$.}			
		\label{fig:H-5}
	\end{center}
\end{figure}

The rotation matrix leading from frame \FII into frame \FB is given by
\begin{equation}
  \mathsf{Q} = \mathsf{R}_3(\varphi_{\mathcal{B}})\mathsf{R}_2(\theta_{\mathcal{B}}-\halfpi)\mathsf{R}_1(\alpha),
\end{equation}
and the transformation of the components of one vector in the body frame \FB to those in the equatorial frame \FII is
\beq
\mathbf{v}_{[\rm II]}= \mathsf{Q}\mathbf{v}_{[\mathcal{B}]}. \label{eq:IItoB}	
\endeq

The spherical triangle shown in Fig. \ref{fig:H-5} allows us to calculate the angle $\alpha$ as a function of the coordinates of the rotation pole and the vertex of the ellipsoid. We have
\beq
\sin\alpha=\sin\theta_\mathcal{P}\sin\Delta\varphi; \qquad  \cos\alpha=\frac{\cos\theta_\mathcal{P}}{\sin\theta_\mathcal{B}},
\endeq
where $\Delta\varphi= \varphi_\mathcal{P} - \varphi_\mathcal{B}$.	

We may recall that the choice of the equatorial and polar vertices $\mathcal{B}$ and $\mathcal{P}$ is such that $(\hat{\bf x}_\mathcal{B}\cdot \mathbf{r}) \ge 0$ and $(\hat{\bf z}_\mathcal{B}\cdot \mathbf{\Omega}) \ge 0$, where $\hat{\bf x}_\mathcal{B}$ and $\hat{\bf z}_\mathcal{B}$ are, respectively, the unit vectors pointing along the $x_\mathcal{B}$- and $z_{\mathcal{B}}$-axes.

The coordinates and other quantities related to the frames are summarized in Table \ref{tab:quant}. 

\begin{table}
	\caption{GLOSSARY}\label{tab:quant}
	\begin{tabular}{ll}
		\hline\\
		&	Quantities related to the frame transformations\\
		\hline\\
		$\mathsf{M}$	& Direction of the companion \\
		${\cal F}_{\rm I}$ & Astrocentric Reference Frame with Inertial Axes\\
		${\cal F}_{\rm II}$ & Equatorial Reference Frame\\
		${\cal F}_{\rm III}$ & Reference Frame attached to the Static Equilibrium Ellipsoid\\
		${\cal F}_{\mathcal{B}}$ & Reference Frame attached to the actual (Dynamic) Equilibrium Ellipsoid\\
		$\mathsf{N}_e$  & Ascending Node of the Orbit in the Equator \\
		$\mathsf{N}$	& Ascending Node of the Orbit in the Fundamental Plane $(x_{\rm I},y_{\rm I})$\\
		$\mathsf{N'}$	& Ascending Node of the Equator in the Fundamental Plane $(x_{\rm I},y_{\rm I})$\\
		$I$ & Orbital Inclination  w.r.t. Fundamental Plane $(x_{\rm I},y_{\rm I})$\\
		$J$& Co-latitude of the Rotation Pole in the Astrocentric Frame ${\cal F}_{\rm I}$ \\
		$\psi$ & Longitude of the Node $\mathsf{N}'$ of the Equator in the Fundamental Plane \\
		$\delta$ & Angle between the radius vector of the companion $\mathsf{M}$ and the major axis of the \\
		&Static Equilibrium Ellipsoid ($x_{\rm III}$-axis) \\
		$\beta$ & Obliquity\\
		$\Omega_{\rm I}$ & Angle between the nodes $\mathsf{N}'$ and $\mathsf{N}$ \\
		$\Omega_{\rm II}$ & Angle between the nodes $\mathsf{N}'$ and $\mathsf{N}_e$ \\
		$\psi+\Omega_{\rm I}$ & Longitude of the Ascending Node of the Orbit
		in the Fundamental Plane  \\
		\hline\\ 
		& Spherical and Rectangular coordinates\\ 
		\hline\\
		$\varphi$ &Longitude of the companion in the equatorial frame ${\cal F}_{\rm II}$ \\
		$\theta$ &Co-latitude of the companion in the equatorial frame ${\cal F}_{\rm II}$ \\
		$\rho$ & Radius vector of a point in the surface of the static equilibrium ellipsoid\\
		$\zeta$ & Radius vector of a point in the surface of the dynamic equilibrium ellipsoid\\
		& (i.e the actual primary body figure)\\
		$x_{\I}, y_{\I}, z_{\I}$ & Coordinates in the Astrocentric reference frame ${\cal F}_{\rm I}$\\
		$x_{\II}, y_{\II}, z_{\II}$ & Coordinates in the Equatorial reference frame ${\cal F}_{\rm II}$\\
		$x_{\III}, y_{\III}, z_{\III}$ & Coordinates in the frame ${\cal F}_{\rm III}$\\
		$x_{\mathcal{B}}, y_{\mathcal{B}}, z_{\mathcal{B}}$ & Coordinates in the frame ${\cal F}_{\mathcal{B}}$\\
		$\varphi_s$ &Longitude of a generic direction in the frame \FII\\
		$\theta_s$ &Co-latitude of a generic direction in the frame \FII \\
		\hline\\
	\end{tabular}
	
\end{table}
\section*{Acknowledgement}
The authors acknowledge the support of FAPESP Proc. 2016/13750-6 ref. Mission PLATO, Proc. 2016/20189-9 (H.A.F.), and 2023/03060-6 (R.A.S.) and CNPq Proc. 303540/2020-6 (S.F.-M.). The authors are thankful to the reviewers for their careful reading of the original version of the paper and their very useful comments.

\end{document}